\documentclass[onecolumn,fleqn,usenatbib]{mnras}

\usepackage{mathptmx}

\usepackage[T1]{fontenc}

\DeclareRobustCommand{\VAN}[3]{#2}
\let\VANthebibliography\thebibliography
\def\thebibliography{\DeclareRobustCommand{\VAN}[3]{##3}\VANthebibliography}

\usepackage{graphicx}	
\usepackage{amsmath}	
\usepackage{amssymb}	
\usepackage{soul}

\usepackage{lineno}

\title[The exoplanetary magnetosphere extension in Sun-like stars]{The exoplanetary magnetosphere extension in Sun-like stars based on the solar wind - solar UV relation}

\author[R. Reda et al.]{
Raffaele Reda,$^{1}$
Luca Giovannelli,$^{1}$\thanks{E-mail: luca.giovannelli@roma2.infn.it (LG)}
Tommaso Alberti,$^{2}$
Francesco Berrilli,$^{1}$
Luca Bertello,$^{3}$
\newauthor
Dario Del Moro,$^{1}$
Maria Pia Di Mauro,$^{2}$
Piermarco Giobbi,$^{1}$
and Valentina Penza$^{1}$
\\
$^{1}$Department of Physics, University of Rome Tor Vergata, Via della Ricerca Scientifica 1, Rome, 00133, Italy\\
$^{2}$INAF - Istituto di Astrofisica e Planetologia Spaziali, Via del Fosso del Cavaliere 100, Rome, 00133, Italy\\
$^{3}$National Solar Observatory, 3665 Discovery Drive, 3rd Floor, Boulder, CO 80303, USA
}

\date{Accepted XXX. Received YYY; in original form ZZZ}

\pubyear{2021}

\begin{document}
\label{firstpage}
\pagerange{\pageref{firstpage}--\pageref{lastpage}}
\maketitle


\begin{abstract}
The Earth's magnetosphere extension is controlled by the solar activity level via solar wind properties. Understanding such a relation in the Solar System is important for predicting also the condition of exoplanetary magnetospheres near Sun-like stars. 
We use measurements of a chromospheric proxy, the Ca II K index, and solar wind OMNI parameters to connect the solar activity variations, on the decennial time scales, to the solar wind properties.
The data span over the time interval 1965-2021, which almost entirely covers the last 5 solar cycles.
Using both cross-correlation and mutual information analysis, a 3.2-year lag of the solar wind speed with respect to the Ca II K index is found. Analogously, a 3.6-year lag is found once considering the dynamic pressure.
A correlation between the solar wind dynamic pressure and the solar UV emission is found and used to derive the Earth's magnetopause standoff distance.
Moreover, the advantage of using a chromospheric proxy, such as the Ca II K index, opens the possibility to extend the relation found for the Sun to Sun-like stars, by linking stellar variability to stellar wind properties.
The model is applied to a sample of Sun-like stars as a case study, where we assume the presence of an Earth-like exoplanet at 1 AU.
Finally, we compare our results with previous estimates of the magnetosphere extension for the same set of Sun-like stars.
\end{abstract}

\begin{keywords}
solar-terrestrial relations -- solar wind -- Sun: UV radiation -- Stars: activity -- Stars: solar-type -- planet–star interactions
\end{keywords}



\section{Introduction}
\label{S-Introduction}

The solar wind is a continuous plasma flow emitted from the upper atmosphere of the Sun, mostly consisting of ions and electrons \citep[e.g.][]{Verscharen19}. At 1 AU, it is characterised by a typical speed ranging between 250 km/s and 800 km/s, a density of a few particles per cubic centimeter, and it carries out a magnetic field of the order of a few nanoTeslas \citep{Parks18}, with a dependency over the solar activity cycle \citep{Poletto2013}. 
In the last 40 years the solar wind has been investigated with increasing deeper details, both in terms of instrument resolution and spacecraft locations.
This refined our understanding of its dynamical properties \citep{Escoubet97,Stone98,Burch16}, opening new insights on the the source regions of the fast solar wind, i.e. the lower solar atmosphere layers of the Coronal Holes \citep[e.g.][]{Bryans2020}.
 It is considered as a natural laboratory to investigate several kinds of processes and mechanisms such as turbulence and intermittency, plasma instabilities, waves and structures, small- vs. large-scale dynamics \citep[e.g.][]{Bavassano1998,Bruno16}. 
Solar wind can be described as a multiscale system whose dynamic occurs over a wide range of scales. 
If we focus on time scales longer than the so-called inertial range (i.e., longer than a few hours), the dynamics of the solar wind is mainly related with solar source mechanisms as active regions, coronal mass ejections, and flares \citep{Tu95}. Thus, these large-scale phenomena are the main reason for the variations of the planetary environments, through the interaction with planetary magnetospheres and/or ionospheres. 
A wide variety of processes are generated, such as geomagnetic storms and substorms, particle precipitation, auroral activity, localized energy transfer processes, impacting the climate and habitability of terrestrial extrasolar planets \citep[e.g.][]{Russell93,Blanc05,Airapetian2020,Galuzzo2021}.
A clear manifestation of the solar wind-magnetosphere interactions is the observed change of the standoff distance of the nose of the magnetospheric cavity, thus affecting both its size and shape. The standoff distance is defined as the distance at which the solar wind dynamic pressure equals the magnetic pressure of the magnetospheric cavity, i.e.,
\begin{linenomath*}
\begin{equation}
    \left| \rho \left( \mathbf v \cdot \nabla \right) \mathbf v \ \right| \simeq \left| -\nabla\left(\frac{B^2}{2\mu_{0}}\right) \right|.
\end{equation}
\end{linenomath*}
Assuming the incompressibility condition for the solar wind, i.e., a constant mass density $\rho = \rho_0$, and writing the magnetic field as a dipolar shape, i.e., $B = \frac{M_E}{r^3}$, being $M_E$ the Earth's dipole moment, we can obtain the standoff distance as
\begin{linenomath*}
\begin{equation}
    R_{MP} = \left(\frac{1}{\mu_0\rho_0}\right)^{1/6} \left(\frac{M_E^2}{v^2}\right)^{1/6}.
    \label{eq:Rs}
\end{equation}
\end{linenomath*}
In the last two decades, several efforts have been made to increase our capabilities in forecasting the dynamical behaviour of the solar wind as well as its effects on Earth \citep{Bothmer07}. Thus, investigating and characterising the relations between long-term solar activity proxies and in-situ solar wind measurements are of crucial impact for any Space Weather and Space Climate forecasting scheme and could be fundamental for characterising Sun-like stars and their interaction with their own planetary systems \citep[e.g.][]{Airapetian2020}. \\
The solar magnetic activity main periodicity is the 22-year Hale cycle \citep{Hale1925}, with the reversal of polarity that results in the well-known 11-year Schwabe solar cycle \citep{Schwabe1844}.
Periods close to 11 years have been found in most solar wind parameters since the very first observations from satellites \citep{Siscoe1978,King1979,Neugebauer1981}. However, a not perfect match of the solar wind long-term behaviour with the shape and phase of the sunspot cycle, stimulated a discussion about the observed lag with geomagnetic indices \citep[see e.g.,][]{Hirshberg73,Intriligator1974, Feynman82}.
As longer time-series of near-Earth solar wind measurements became available, several studies have investigated the periodicity of the solar wind parameters and the relation between the Sunspot Number and solar wind proxies \citep{Petrinec1991, Kohnlein96,ElBorie02,Katsavrias12, Richardson12, Li2016,Li2017, Venzmer2017, Samsonov19}, or geomagnetic data such as the aa index \citep{Echer2004,Dmitriev2005,Lockwood2009, Du2011}.
\begin{figure*}
    \centering
    \includegraphics[scale=0.65]{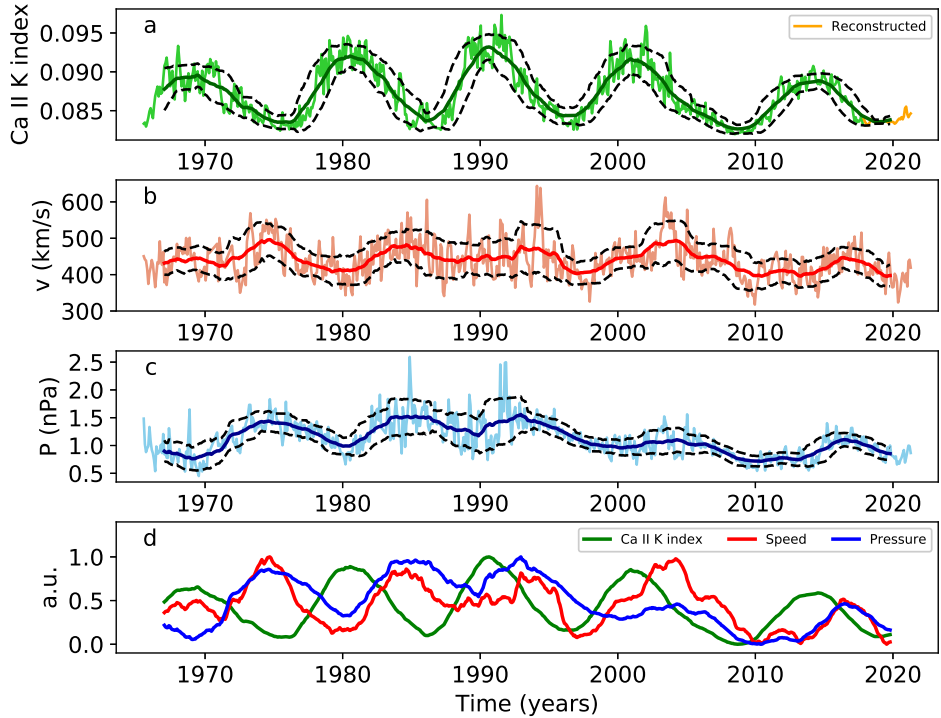}
    \caption{Monthly means (light colours) and superimposed 37-month moving averages (dark colours) of: a) Ca II K index, b) Solar Wind speed and c) Solar Wind dynamic pressure. In the upper panel, the orange line shows the reconstructed Ca II K index obtained using Mg II index. d) Comparison of Ca II K index (green), SW speed (red) and SW dynamic pressure (blue) normalized between 0 and 1. In panels a, b, and c the dashed lines indicate the 1-sigma confidence interval.}
    \label{Ca-v-p}
\end{figure*}
However, this is the first study to analyse the near-Earth solar wind measurements in relation to Ca II K index over the last five solar cycles. The emission in the Ca II K resonance line, related to the mean emission of the Sun's chromosphere, has been proven to be a great proxy of solar activity \citep[see e.g.,][]{Judge2006, 2016SoPh..291.2967B, Chatzistergos2019}.\\
Starting from the 1960s, astronomers began looking with particular attention at other "suns" (i.e., stars with physical properties similar to solar ones or Sun-like stars), with the aim to search for stellar cycles in solar analogs and to understand where the Sun stands on a broader context \citep{Wilson68}. To carry out this purpose, a long-term observational campaign regarding their emissions in Ca II H \& K lines, expressed in terms of the dimensionless S-index \citep{Wilson78, Vaughan78}, was conducted at Mount Wilson Observatory (MWO) starting from 1966 \citep{Wilson68} and then continued at Lowell Observatory starting from 1995 \citep[see e.g.,][]{Hall98}. The data provided by those surveys have enormously improved our knowledge about the long-term chromospheric variations of Sun-like stars, as well as how these variations are connected with changes in brightness \citep[see e.g.,][]{Skumanich72, Baliunas95, Radick98}. \\
When we study such type of stars, the Sun represents a sort of Rosetta stone, so that our understanding about how its magnetic activity affects the solar-system planets can be seen as a starting point to assess the way Sun-like stars influence the environment around them.
Assessing the magnetospheric compression of planets orbiting Sun-like stars is a fundamental point to evaluate their habitability conditions \citep{Airapetian2020}. It is well known that the presence of a large enough magnetosphere is a crucial point to shield the planetary atmosphere from the effects of stellar activity, since sufficiently strong stellar wind may also be able to erode it \citep[see e.g.,][]{Lammer2012, Cohen2015, Rodriguez-Mozos2019}. \\
The goal of this work is to extend the model of the Earth's magnetosphere standoff distance based on the Ca II K index to exoplanets orbiting Sun-like stars. To achieve this goal, we take advantage of chromospheric measurements provided by the MWO campaign, by selecting a sample of Sun-like stars that are in a faculae-dominated activity regime as the Sun \citep[see e.g.,][]{Radick98, Shapiro2016}. Such selection has been made by considering only Sun-like stars with a Rossby number $R_{0} > 1$ which, as pointed out by \citet{Reinhold19}, corresponds to a stellar age $\gtrsim 2.55$ Gyr and to the transition to faculae-dominated activity.
%


\section{Datasets description}
\label{S-Datasets}

To study the relation between solar magnetic variability and near-Earth solar wind parameters, on decennial time-scales, we need to use datasets which cover a sufficiently extended time interval.
As previously described, the magnetic activity of the Sun can be quantified by using different solar indices \citep{Hathaway2010}.
For this work we use a physical index which measures the mean properties of the solar chromospheric emission: the Ca II K index. Monthly measurements of this index starting from 1907 and covering more than one century are publicly available from the National Solar Observatory (NSO) \footnote{\url{https://solis.nso.edu/0/iss/}}.
The Ca II K index dataset contains measurements up to October 2017, but it is possible to use other solar activity proxies linked to chromospheric emission, such as the Mg II index \citep{Viereck01}, to extend the analysis almost up to date.\\
For the solar wind we use data available from the OMNI database, which provides various near-Earth solar wind parameters at different time resolutions (more details about the data are given in Sect. \ref{OMNI data}). In particular, we focus our attention to the hourly-resolution measurements of the plasma ion density $n$ and speed $v$, thus determining the solar wind dynamic pressure $P$, defined as $1/2\,m_{p} \, n \, v^{2}$, where we assume proton mass ($m_p$) as the mean ion mass. The OMNI database provides plasma measurements only starting from 1965 and, therefore, covers a shorter time interval with respect to the Ca II K index time series. This places a limit on the length of the time period over which a relation between solar proxies and solar wind parameters can be studied. Despite this, by using the Mg II index to reconstruct Ca II K index to date, the latter and the solar wind parameters time series have an overlapping time period which goes from July 1965 up to April 2021, covering almost entirely the last five solar cycles (SCs 20-24). \\
The properties of each dataset are described in detail later in this section. Here we concentrate on the first steps in the data treatment for the OMNI solar wind dataset. We first bin our data to obtain monthly means.
Therefore, starting from the monthly values of Ca II K index, solar wind speed and dynamic pressure, we follow the approach used by \citet{Kohnlein96} and apply a 37-month moving average to look at the long-term behaviour of these quantities. These time series are shown in Figure \ref{Ca-v-p}. The size of the time window used for the moving average allows to filter and remove the effects of the solar variability related to the solar rotation, transient phenomena, and any other source of variation below the yearly time-scales.\\
The availability of the observations in the Ca II H \& K lines provided by the HK Project at the Mount Wilson Observatory \citep{Wilson68, Wilson78, Duncan91, Baliunas95} is fundamental to extend the model calibrated on the Sun-Earth system to Sun-like star systems.
These measurements are accessible for thousands of stars and constitute a broad and long dataset spanning nearly 30 years in many cases. We take advantage of the availability of such measurements to relate, in stars other than the Sun, the mean chromospheric emission to stellar wind dynamic pressure, and hence to study its impact on the exo-planetary magnetospheres. \\
In the following subsections a detailed description of the datasets used for this work will be provided.

\subsection{Solar wind OMNI dataset}
\label{OMNI data}
We used solar wind data from the OMNI dataset at 1-hour resolution freely retrieved at \url{https://cdaweb.gsfc.nasa.gov/cgi-bin/eval1.cgi}. This dataset consists of a collection of solar wind magnetic field and plasma parameter data coming from several spacecrafts located near the L1 Lagrangian point at a distance of $\sim$200 Earth radii. The measurements taken at L1 are then shifted to the nose of the bow shock ($\sim$14 Earth radii) by considering several factors such as the geometry of the Earth-spacecraft separation vector, the shape and the orientation of the solar wind variation phase front and the direction of the solar wind flow \citep{Weimer02,Weimer03}. By assuming that solar wind parameter values lie on a planar surface (i.e., the phase front) convected by the solar wind, we are able to propagate what is observed at the L1 point to a different place at the time that the phase front sweeps over that location \citep{Weimer08}. The family of spacecraft considered for building up the OMNI database is formed by IMP, ISEE, ACE, Wind, and Geotail \citep{King05}, thus allowing to cover the period from 1965 to date \citep{Richardson01}. 

\subsection{Ca II index and Mg II index datasets}
\label{Ca II K dataset}
The Ca II K 0.1 nm emission index data are derived from the series of spectroheliograms taken at Kodaikanal Solar Observatory (India, 1907 - 2013), from the K-line monitor program of disk-integrated measurements of the National Solar Observatory (NSO) at Sacramento Peak (USA, 1976 - 2015), and from the Integrated Sunlight Spectrometer  on the Synoptic Optical Long-term Investigations of the Sun (SOLIS) telescope managed by NSO (USA, 2006 - 2017). Since October 2017 the SOLIS facility has been offline, pending its relocation to a permanent site at Big Bear Solar Observatory (California, USA). The procedure to combine these three datasets into a single disk-integrated Ca II K 0.1 nm emission index time series composite is described in \cite{2016SoPh..291.2967B}. The composite is created by cross-calibrating overlap periods between these three sets of observations, with the SOLIS time series used as a fiducial. During the period 2007 to 2013 observations were taken both at Sacramento Peak and by the SOLIS instrument. This time-span is long enough to cross-calibrate the two sets of measurements, since both time series are strongly modulated by the 11-year solar cycle. A scaling factor of 1.032 was derived to put the Sacramento Peak values into the same scale as the SOLIS measurements. A similar approach is then applied to rescale the Ca II K index values derived from the Kodaikanal observations, using 21 years (1978 to 1999) of overlapping observations with Sacramento Peak. For more details, the reader can refer to \citet{2016SoPh..291.2967B}. \\
As previously mentioned, monthly values of the Ca II K index are not available after October 2017. This lack of data can be overcome by using other physical indices related to the chromospheric emission of the Sun, whose measurements are available to date. To this scope, we use the Mg II composite from the University of Bremen, which is derived from combining several satellite instruments \citep{Viereck04} and has been proven to be an excellent proxy for the solar UV irradiance \citep{Wit2009}  related to the interaction with the Earth's high atmosphere \citep[see e.g.,][]{Larkin2000} and thermosphere \citep{Bigazzi2020}.
The Mg II index, defined as the core-to-wing ratio of the Mg II doublet centred at 280 nm, has been measured from November 1978, and it is freely accessible with daily resolution \footnote{\url{http://www.iup.uni-bremen.de/UVSAT/Datasets/mgii}}. Starting from the daily values, we calculate the monthly means of Mg II index and we note that the latter strongly correlate with Ca II K index (r = 0.95) in the time interval November 1978 - October 2017. Then, by using the linear relation $\mathrm{Ca\;II\,K} = 0.5619\; \mathrm{Mg\;II} - 0.0014$,  we extend the monthly dataset of Ca II K index until April 2021.

\subsection{Mount Wilson Observatory dataset}
\label{MWO data}
The first long-term observational campaign to study and characterise the magnetic activity behaviour of stars other than the Sun, named HK Project, has been conducted at the Mount Wilson Observatory. In order to search for stellar analogues to the solar cycle, the emission in the chromospheric H (393.4 nm) and K (396.8 nm) lines of the Ca II has been monitored from 1966 to 1995 for thousands of stars \citep{Wilson68, Wilson78, Duncan91, Baliunas95}. The measurements from the MWO are expressed in term of the S-index, a dimensionless quantity which is defined as the ratio of emission in the Ca II H \& K line cores to that in two nearby reference bandpasses (see the definition provided by \citet{Vaughan78} for further details). 
The MWO S-index is the most used stellar magnetic activity index and, unlike the Ca II K 0.1 nm emission index described in Sect. \ref{Ca II K dataset}, quantifies the chromospheric emission of a star accounting for the flux of emission in both Ca II H \& K lines. Obviously, the Ca II K index and the S-index are strictly related. Several authors have investigated the relationship between these two indices, in order to place the Sun in the stellar S-index scale and making it possible to compare the solar activity to that of Sun-like stars \citep{Duncan91, White1992, Baliunas95, Radick98, Hall2004}. One of the most recent works to assess the relationship between Ca II K index and S-index has been done by \citet{Egeland17}, providing the following linear relation:

\begin{equation}
\label{S-CaII relation}
\mathrm{S\;(Ca\;II\;K)} = (1.50 \pm 0.13)\; \mathrm{Ca\;II\,K} + (0.031 \pm 0.013)
\end{equation}
This relation allows to accurately place the Sun on the stellar S-index scale, providing a simple procedure to switch from one index to the other. \\
The observations from the MWO have constituted, during the last decades, an important basis for studying processes analogous to solar activity and cycle, as well as how they are related to stellar properties \citep{Vaughan80, Durney81, Baliunas95, Saar99, Hall08, Olah16}. Moreover, they have allowed to study the way chromospheric variability is connected with changes in brightness, whose phase difference reveals the stellar activity regime, spot-dominated (anti-phase) or faculae-dominated (in phase) \citep{Radick98, Reinhold19}. 
Recently, the Mount Wilson HK Project data for almost 2300 stars have been released by the National Solar Observatory (NSO) \footnote{\url{https://nso.edu/data/historical-data/mount-wilson-observatory-hk-project/}}.


\section{Long-term correlations of solar wind parameters and solar activity}
\label{S-analysis}

\begin{figure*}
    \centering
    \includegraphics[scale=0.5]{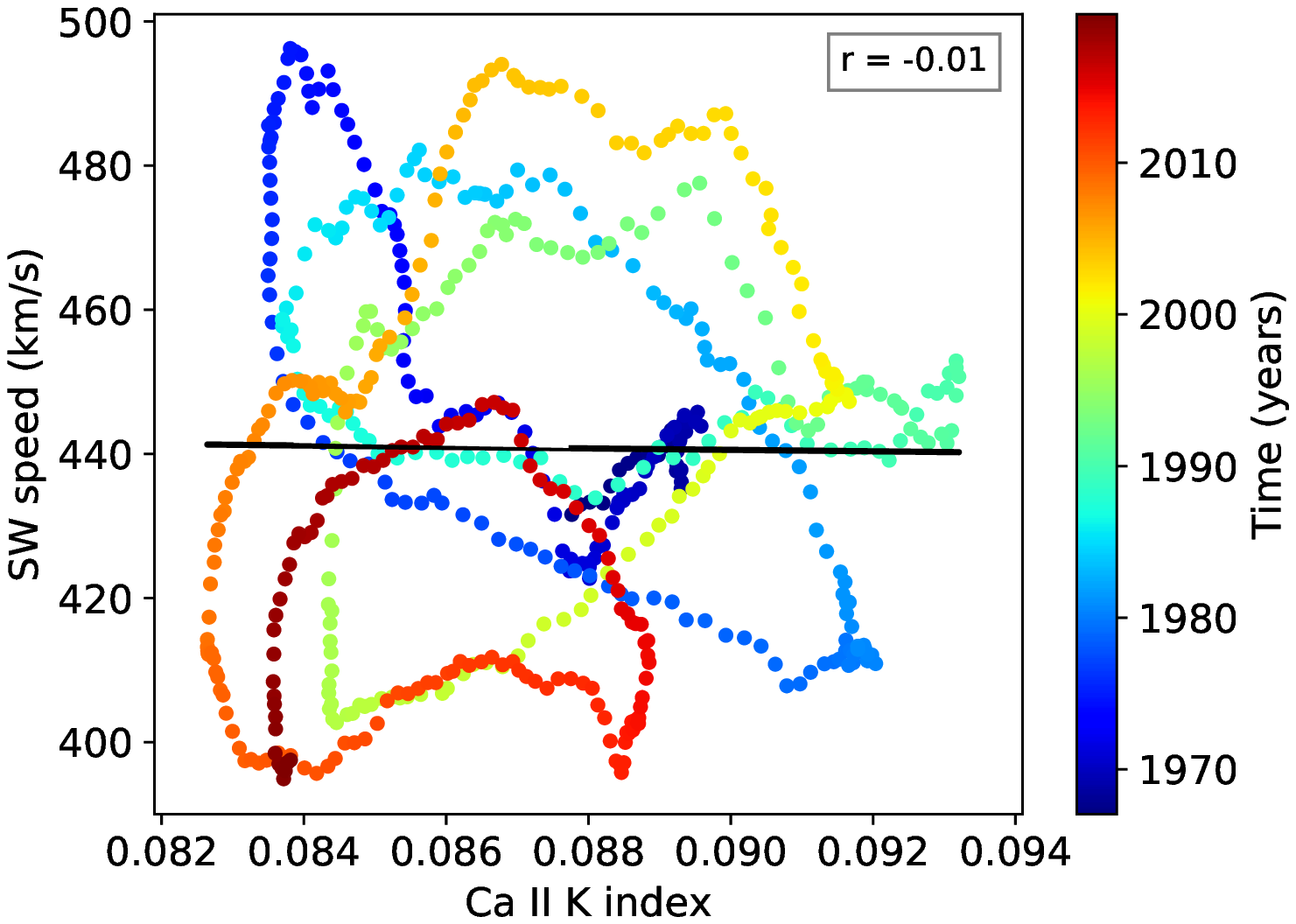}
    \includegraphics[scale=0.5]{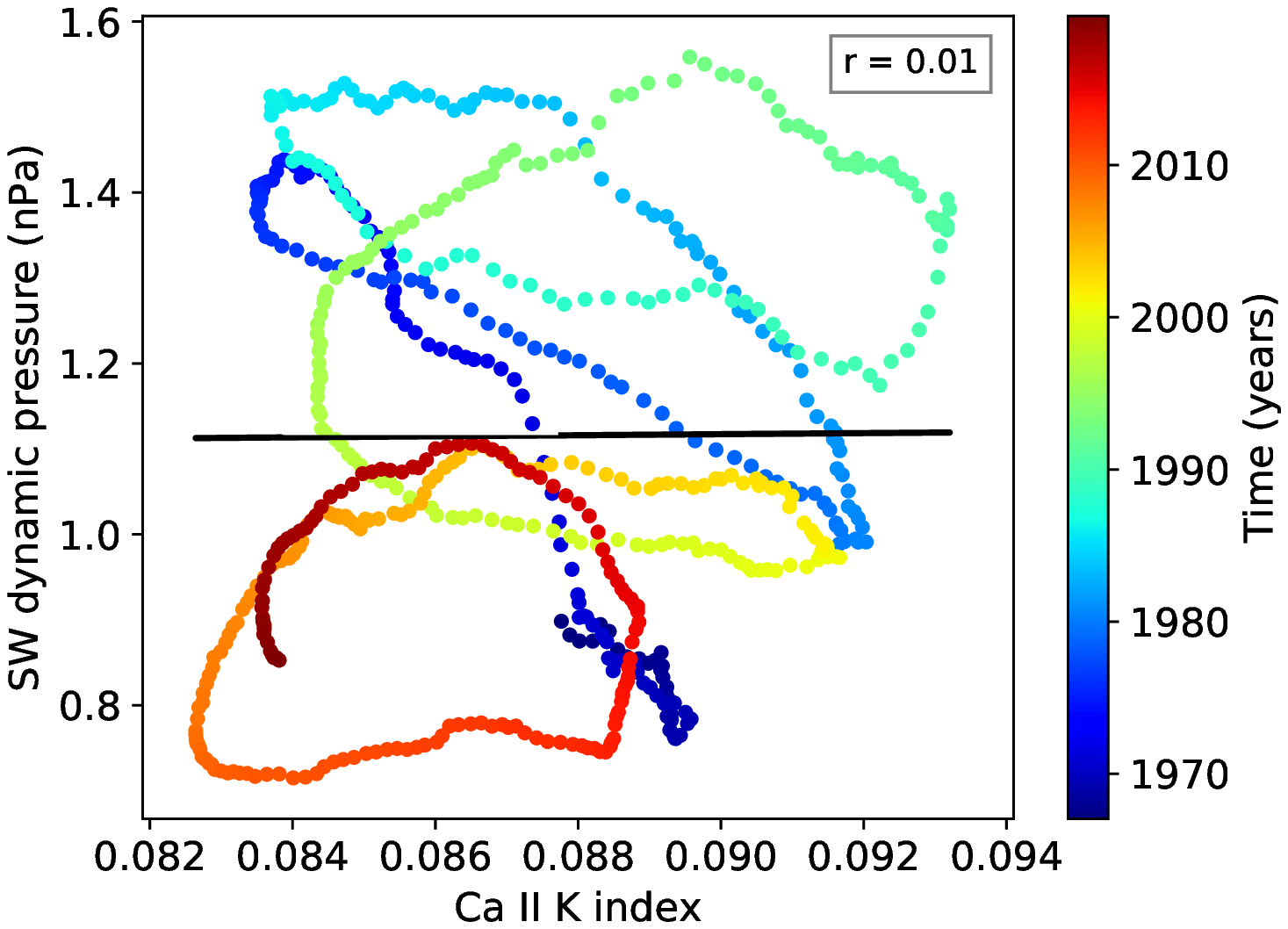}
    \caption{Scatter plot of 37-month moving averages of Ca II K index and, respectively, solar wind speed (left panel) and solar wind dynamic pressure (right panel). The time is represented by the color map. The correlation coefficient is, respectively, -0.01 and 0.01.}
    \label{Scatter color map ca-v}
\end{figure*}

\begin{figure*}
    \centering
    \includegraphics[scale=0.7]{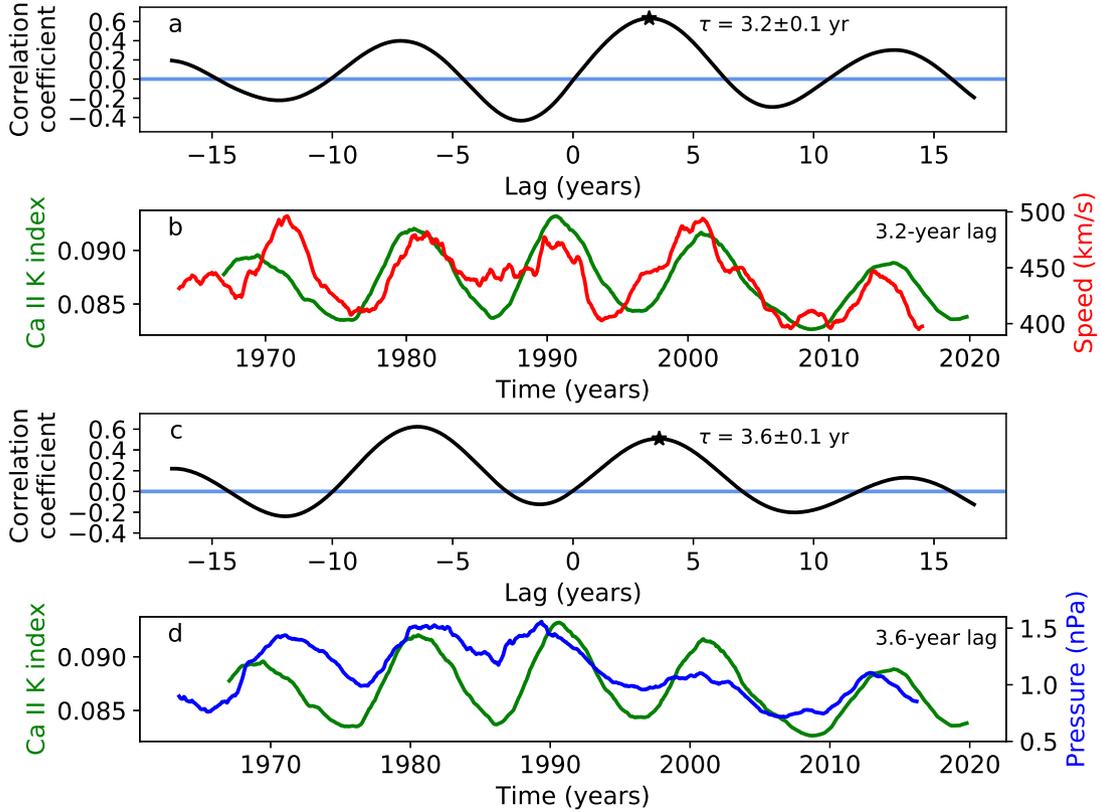}
    \caption{a) Cross-correlation between Ca II K index and Solar Wind speed. b) Solar Wind speed shifted backward with respect to Ca II K index by 3.2 years (time lag corresponding to maximum amplitude of the cross-correlation). c) Cross-correlation between Ca II K index and Solar Wind dynamic pressure. d) Solar Wind pressure shifted backward with respect to Ca II K index by 3.6 years (time lag corresponding to maximum amplitude of the cross-correlation). }
    \label{cross correlations}
\end{figure*}

As a first step in our analysis, we assess the relationship between solar activity and solar wind parameters by computing the Pearson's correlation coefficient over the whole time extent of the dataset. Our analysis is applied to the 37-month moving average quantities presented in Sect. \ref{S-Datasets}. As shown in the scatter plots of Figure \ref{Scatter color map ca-v}, we found an almost zero correlation coefficient (r = -0.01) between Ca II K index and solar wind speed, with a similar result also for the dynamic pressure (r = 0.01). Hence, for the whole time interval, from July 1965 to April 2021, we do not find a significant correlation (the p-value is higher than 0.05) between Ca II K index and the solar wind parameters.
As shown in previous studies, the declining phase of solar cycles are characterised by the presence of high-speed solar wind streams \citep{Gosling76,Luhmann09, Tokumaru10, Richardson12}.
Thus, it is reasonable to expect a time lag between solar activity and solar wind response. To investigate this hypothesis, we compute the cross-correlation between Ca II K index and solar wind speed, where we assume that the latter has a delayed response to changes in solar activity, which means that we are considering only positive time lags of the solar wind with respect to Ca II K index (the result for negative lags is also shown for completeness). As shown in the top panel of Figure \ref{cross correlations}, the correlation coefficient between the two quantities peaks, with a value of 0.65, at time lag of $3.2\pm0.1$ yr. 
The two time-series as visible in the panel b of Figure \ref{cross correlations}, where the solar wind speed has been back-shifted by the time lag found, have a quite similar phase when a time delay is considered. This result is quite in agreement to that reported by \citet{Li2016}, which found that the daily means of the solar wind velocity lag the ones of SSN by about 3 yr (see their fig. 3). A 3-year time shift was also found by \citet{Venzmer2017} for the correlation of the yearly averages of the same quantities. \\
A slightly bigger time lag has been found performing the cross-correlation of Ca II K index with the solar wind dynamic pressure data. In this case, we find a moderate positive correlation with a correlation coefficient of 0.57 for a time lag of $3.6\pm0.1$ {yr}, as shown in the panel c of Figure \ref{cross correlations}. Although this latter value is lower than the one found for the solar wind speed, it is higher than the aforementioned correlation coefficients found in literature for SSN and solar wind velocity (0.2 in \citealt{Li2016} and 0.45 in \citealt{Venzmer2017}). Further comments on the strength of the correlation can be found in the discussion section.

To assess our results in a stronger framework, we also explored the nonlinear features of shared information between the Ca II K index and solar wind parameters using the mutual information analysis~\citep{Shannon48}. Given a pair of time series $(x(t_j),y(t_k))$ the mutual information coefficient ($MI$) is defined as
\begin{linenomath*}
\begin{equation}
MI = \sum_{j, k} p(x(t_j), y(t_k)) \log\frac{p(x(t_j), y(t_k))}{p(x(t_j)) p(y(t_k))}
\end{equation}
\end{linenomath*}

where $p(x,y)$ is the joint probability of observing the pair of values $(x,y)$, while $p(x)$ and $p(y)$ are the independent distributions. For statistically independent time series $MI = 0$, while for correlated time series $MI \ge MI_{th}$, where $MI_{th}$ is a threshold associated with a particular statistical significance level. For the following analysis we adopt a threshold value $MI_{th}$ that corresponds to a $95\%$ statistical significance.
\begin{figure}
   \centering
    \includegraphics[scale=0.5]{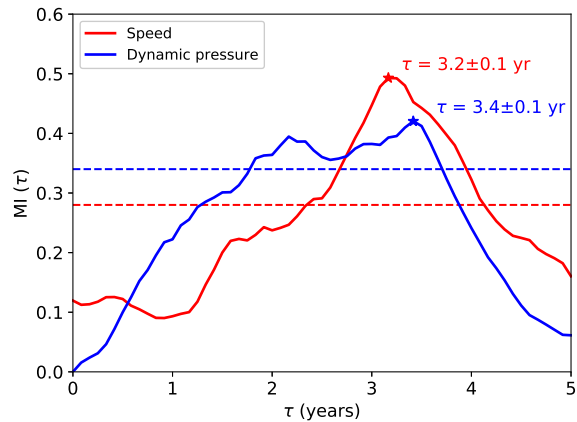}
    \caption{The mutual information coefficient shared between the Ca II K index and the solar wind dynamic pressure (blue line) and speed (red line). The horizontal dashed lines refer to the 95\% statistical significance level.}
    \label{MI}
\end{figure}

As reported in Figure \ref{MI} the mutual information coefficient reaches its maximum for time delays of $\sim 3.2$ yr and $\sim 3.4$ yr for the solar wind speed and dynamic pressure with respect to the Ca II K index, respectively. 
These values are consistent with those estimated via the cross-correlation analysis (see Figure \ref{cross correlations}).
This seems to suggest that there is a significant non-null probability of observing a relation between the Ca II K index and solar wind parameters. In the following, since the same delay is observed both for the MI and the cross-correlation analysis, we use the results from the latter.

\begin{figure*}
    \centering
    \includegraphics[scale=0.5]{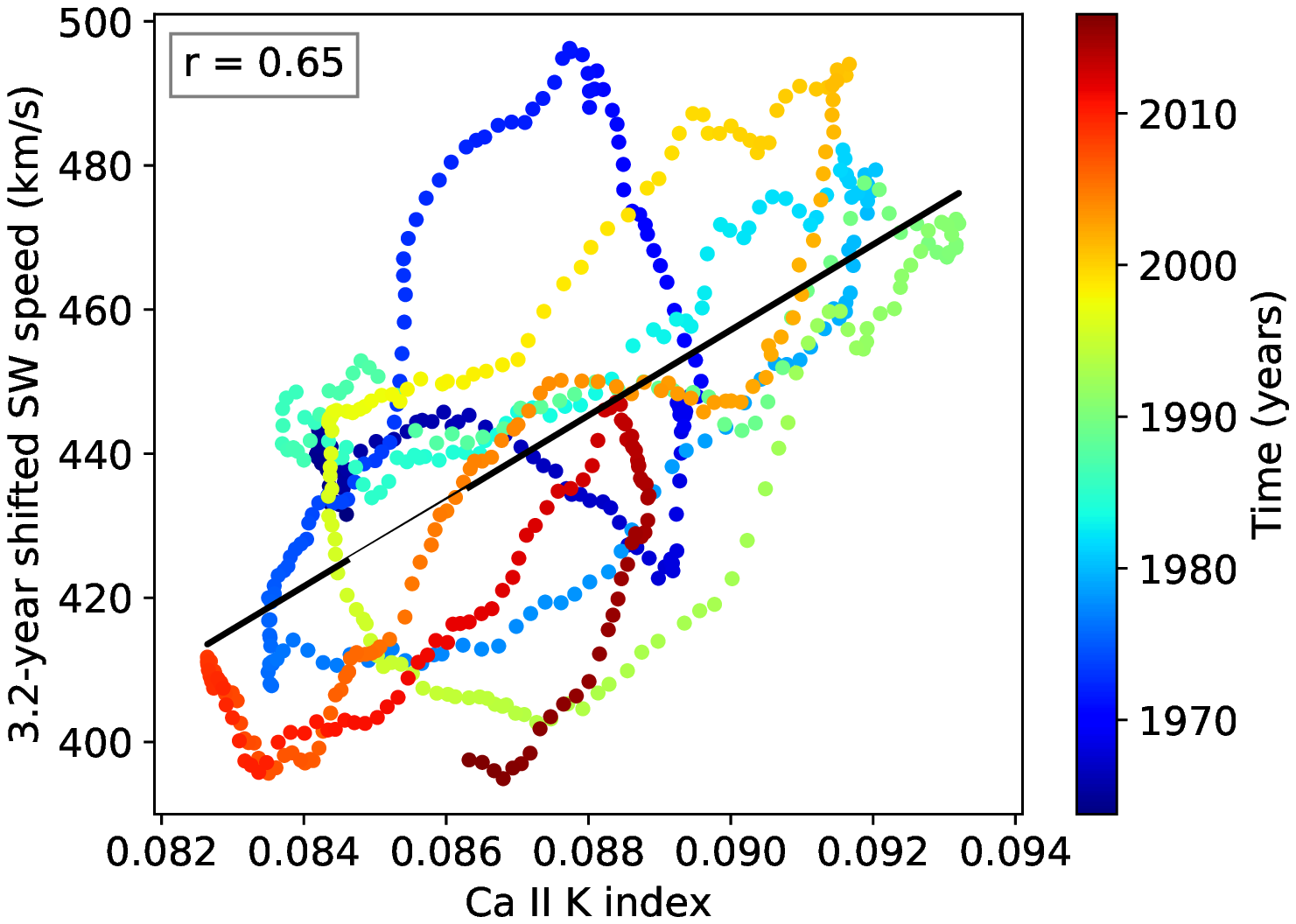}
    \includegraphics[scale=0.5]{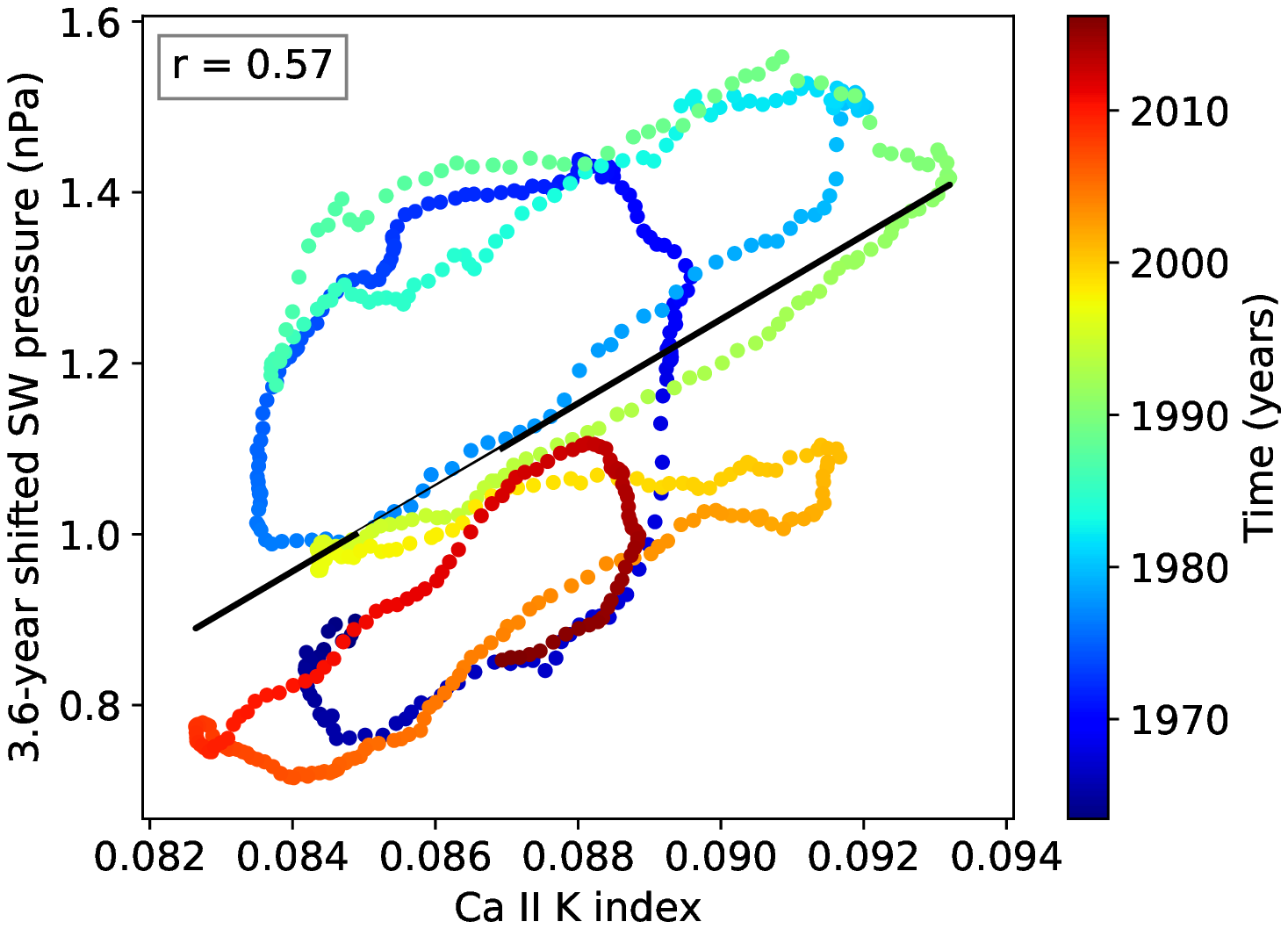}
    \caption{Scatter plot showing the relation of Ca II K index with solar wind speed (left panel) and the solar wind dynamic pressure (right panel) once the lags from cross-correlation analysis have been considered. The correlation coefficient is, respectively, 0.65 and 0.57. The color map shows the evolution of the relation over the time.}
    \label{Shifted scatter plot}
\end{figure*}

By using the above time lags, we found linear relationships for the Ca II K index with both solar wind speed and dynamic pressure. The relationships between these quantities are shown in the two scatter plots in Figure  \ref{Shifted scatter plot}, where the two black lines show the best linear fits to the data points. The corresponding empirical equations are:
\begin{linenomath*}
\begin{equation}
v\,(\mathrm{km/s}) = (5930 \pm 280)\;\mathrm{Ca\;II\,K} - (76 \pm 24).
\label{fit_speed}
\end{equation}
\end{linenomath*}
\begin{linenomath*}
\begin{equation}
P\,(\mathrm{nPa}) = (49.1 \pm 2.8)\;\mathrm{Ca\;II\,K} - (3.17 \pm 0.24);
\label{fit_pressure}
\end{equation}
\end{linenomath*}
We can conclude that, by considering a time lag, the correlation coefficients between Ca II K index and the two solar wind parameters investigated here significantly increase. Similar results were recently reported by \cite{Samsonov19} using yearly averaged data for the last five solar cycles. They found that the maximum correlation of both solar wind speed and dynamic pressure with the SSN is reached with a 3-year time lag. In particular, they found a correlation coefficient r = 0.57 between SSN and SW dynamic pressure, in line with our results of r =  0.57 for 3.6-year lag. Moreover, they found r = 0.68 focusing only over the last three solar cycles taking 2-year lag. If we limit our analysis on the last three solar cycles as well, we find the same correlation coefficient (r = 0.68) for 2.4-year lag.\\

\section{Magnetosphere extension}
\label{S-Magnetosphere}
The magnetosphere shields the Earth's atmosphere from the erosion processes due to solar wind, energetic particles, and radiation. Thanks to the magnetic shielding, the Earth has retained its atmosphere for a long time, but this is not the case of all Solar-system planets.
Because of the lack of a strong intrinsic magnetic field, Mars has suffered a significant solar wind-induced atmospheric loss \citep[see e.g.,][]{Kass1995, Chassefiere2004, Dong2018}, with the result of nowadays very thin atmosphere. \\
Since the Earth's magnetopause standoff distance is mainly controlled by the solar activity via the solar wind pressure, it is interesting to see how its average value changes over a solar cycle. Moreover, assessing the extent of a planetary magnetosphere constitutes a point of fundamental importance also in the characterisation of extra-solar planets, as the presence of a large enough magnetic shield  protecting the underlying atmosphere is directly related to any habitability evaluation.

\subsection{Earth's magnetosphere}
In the previous section we found a relation which allows to connect the 37-month moving averages of a solar UV proxy, the Ca II K index, to the solar wind dynamic pressure. Once the latter is known, the size of the Earth's magnetosphere on the day-side can be calculated by balancing the planetary magnetic pressure with the solar wind dynamic ram pressure. Starting from the relation provided by Eq. 22 in \cite{Greissmeier04}, we introduce the following equation in which the wind dynamic pressure is replaced by the Ca II K index:
\begin{linenomath*}
\begin{equation}
\label{Eq_magnetosphere}
    R_{MP} = \left[\frac{\mu_{0}\,f_{0}^{2}\,M_{E}^{2}}{8\pi^{2}10^{-9}(\alpha\;Ca\;II\,K + \beta)}\right]^{1/6}
\end{equation}
\end{linenomath*}
where $\alpha=49.14$ and $\beta=-3.17$ are the parameters of the linear regression from the previous section, $\mu_{0}$ is the vacuum permeability, $M_E$ is the Earth's magnetic moment, while $f_0$ is a form factor to take into account for the non-spherical shape of the Earth's magnetosphere. For the latter parameter we assume the value $f_{0}=1.16$, as in \citet{See2014}. Regarding the Earth's magnetic moment, even if its value has ranged from $M_{E}=5\cdot10^{22}\mathrm{Am^{2}}$ to $M_{E}=11\cdot10^{22}\mathrm{Am^{2}}$ during the last 12 kyr \citep[see e.g., the review by][]{Olson2006}, considering the time-scales investigated in this work it can be assumed to be constant. For this reason, here we use the value $M_{E}=8\cdot10^{22}\mathrm{Am^{2}}$, as in \citet{See2014}.
\begin{figure}
    \centering
    \includegraphics[scale=0.45]{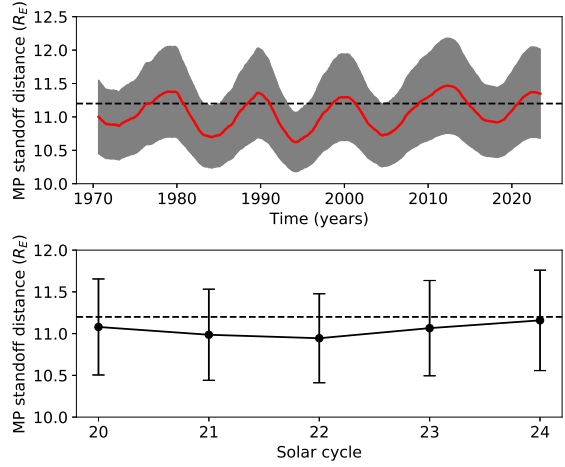}
    \caption{Earth's magnetopause standoff distance according to Eq. \ref{Eq_magnetosphere} for the time interval 1970-2023 (upper panel) and average values for the last 5 solar cycles 20-24 (lower panel). The confidence interval is shown with the shaded gray area in the upper panel and with error bars in the lower one. In both panels the dashed line shows, as reference value, the average standoff distance of the magnetopause \citep{Shue97}.}
    \label{MP_standoff}
\end{figure}
Figure \ref{MP_standoff} shows the Earth's magnetopause standoff distance computed by using Eq. \ref{Eq_magnetosphere} for two cases: 37-month moving average time series by taking into account for the 3.6-year lag of the solar wind dynamic pressure with respect to Ca II K index, as found with the cross-correlation analysis, and average values for each solar cycle. The confidence intervals have been estimated considering the errors of the fit parameters in Eq. \ref{fit_speed} and \ref{fit_pressure}. The magnetopause standoff distance, computed by using the Eq.
12 in \citet{Shue97} and assuming the mean $B_z$ and $P$ values from the OMNI dataset, is shown for comparison in both panels. \\
The presence of the time delay makes possible to compute the standoff distance of the magnetopause up to 2023, as shown in the upper panel of Figure \ref{MP_standoff}, allowing  to forecast future trend.
We estimate that the average extension related to solar cycle 24 should peaks in mid-2022. \\

\subsection{Extending the model to exoplanets orbiting Sun-like stars}
The relation provided by Eq. \ref{Eq_magnetosphere} allows to use a chromospheric proxy of the solar activity to estimate the Earth's magnetopause standoff distance. As the Ca II K index is a physical proxy in principle measurable in each star, such relation can be also extended to stars with properties similar to the Sun. Hence, this relation is very useful, and it can be employed to study the effect of stellar winds on the magnetosphere of Earth-like planets orbiting stars for which similar measurements are available. To this scope a wide dataset has been provided by the HK Project at the Mount Wilson Observatory, where the emissions in the Ca II H \& K lines, expressed in term of the S-index, have been monitored for a broad sample of stars and for long time intervals (up to 30 years for some stars). The relationship between the Ca II K index and the S-index has been already described in Sect. \ref{MWO data}, where we also introduced the formula which allows to switch from the former to the latter index.
\begin{table*}
    \centering
    \begin{tabular}{lccccccc}
        \hline
        Star & Spectral type & $T_{eff}$ & log g & log $R_{0}$ & Ca II K & $R_{MP}$ (this work) & $R_{MP}$ (\citealt{See2014})\\
        &  & ($K$) & ($cm\,s^{-2}$) &  &  & ($R_E$) & ($R_E$) \\
        \hline
        \hline
        Sun & G2 V & $5770\pm18$ & $4.44\pm0.06$ & $+0.307\,^{+0.017}_{-0.013}$ & $0.087\pm0.003$ & $11.0\pm0.6$ & 10.2
        \\
        \hline
        HD 10780 & K0 V & $5327\pm44$ & $4.54\pm0.06$ & $+0.124\,^{+0.000}_{-0.000}$ & $0.164\pm0.017$ & $8.63\pm0.29$ & $8.83$ \\
        HD 100180 & G0 V & $5989\pm44$ & $4.38\pm0.06$ & $+0.290\,^{+0.000}_{-0.000}$ & $0.089\pm0.012$ & $10.91\pm1.03$ & $10.74$ \\
        HD 13043 & G2 V & $5897\pm44$ & $4.27\pm0.06$ & $+0.324\,^{+0.000}_{-0.004}$ & $0.078\pm0.011$ & $12.05\pm1.91$ & $10.06$ \\
        HD 179958 & G4 V & $5760\pm44$ & $4.39\pm0.06$ & $+0.324\,^{+0.016}_{-0.017}$ & $0.080\pm0.011$ & $11.77\pm1.63$ & $11.02/10.59$ \\
        HD 185144 & G9 V & $5246\pm44$ & $4.55\pm0.06$ & $+0.253\,^{+0.006}_{-0.000}$ & $0.125\pm0.014$ & $9.38\pm0.42$ & $9.75/9.50$ \\
        HD 34411 & G1.5 V & $5911\pm44$ & $4.37\pm0.06$ & $+0.347\,^{+0.534}_{-0.119}$ & $0.076\pm0.011$ & $12.37\pm2.29$ & $10.8$ \\
        HD 71148 & G5 V & $5818\pm44$ & $4.29\pm0.06$ & $+0.290\,^{+0.009}_{-0.010}$ & $0.082\pm0.011$ & $11.54\pm1.42$ & $10.56/10.15$ \\
        HD 76151 & G3 V & $5790\pm44$ & $4.55\pm0.06$ & $+0.169\,^{+0.008}_{-0.000}$ & $0.137\pm0.015$ & $9.10\pm0.37$ & $10.07/9.30$ \\ 
        HD 86728 & G3 V & $5700\pm44$ & $4.29\pm0.06$ & $+0.340\,^{+0.004}_{-0.000}$ & $0.076\pm0.011$ & $12.37\pm2.29$ & $10.83$ \\
        HD 9562 & G1 V & $5939\pm44$ & $4.13\pm0.06$ & $+0.390\,^{+0.004}_{-0.000}$ & $0.071\pm0.011$ & $13.61\pm4.44$ & $10.74$ \\
        
        \hline
    \end{tabular}
    \caption{Colum 1: star ID; Column 2: spectral type according to SIMBAD; Columns 3 and 4: effective temperature and surface gravity from \citet{Valenti05}; Column 5: logarithm of the Rossby number ($R_0$) from Table 1 in \citet{Marsden14}; Column 6: average Ca II K index value as obtained by the MWO measurements using Eq. \ref{S-CaII relation}. For the Sun case, the reported Ca II K index is the mean value of the dataset described in Sect. \ref{Ca II K dataset}. The last two columns show the comparison of the magnetospheric standoff distances from this work and from \citet{See2014} for fictitious Earth-twin planets orbiting these stars. In the last column, stars with large activity ranges are listed with minimum and maximum standoff distances.}
    \label{tab_magnetosphere}
\end{table*}
To test our relation for stars other than the Sun, we selected a set of ten Sun-like stars which fulfil two conditions: observations in the Ca II H \& K lines are available from Mount Wilson for at least one full UV stellar cycle; they are characterised by a Rossby number $R_{0} > 1$ which indicates that the star is in a faculae-dominated activity regime, like the Sun \citep{Reinhold19}.
The spectroscopic parameters of these stars \citep{Valenti05} and the stellar Rossby number \citep{Marsden14} are listed in Table \ref{tab_magnetosphere}. 
In order to apply our relation, we firstly computed the mean stellar S-index value and then, by using Eq. \ref{S-CaII relation}, for each star we calculated the Ca II K index from the S-index. Finally, by using Eq. \ref{Eq_magnetosphere} we computed the expected mean magnetopause standoff distance for hypothetical Earth-twin exoplanets orbiting at 1 AU around the host stars. 
As we have no direct measurements of stellar wind properties, assessing the lag between these and the UV stellar cycle is a difficult task. 
Therefore, using our model is only possible to determine the mean stellar wind properties. This is trivial in the case of almost constant UV emission, whereas in the case of stars showing a cyclic behaviour, it is necessary to measure at least a full cycle. Nevertheless, in this study we concentrate on a short sample of stars similar to the Sun and showing a UV cycle. The stars selected in this study have been observed for at least a full UV stellar cycle, thus the mean S-index is a robust estimate of the magnetopause standoff distance mean value. We plan to provide a more detailed analysis on a wider set of stars in an upcoming study that will include a discussion on the amplitude of the UV stellar cycle and the related stellar wind properties.   
The results of the present analysis can be found in the 7th column of Table \ref{tab_magnetosphere}, where we compared the magnetosphere sizes from our relation to that obtained by \cite{See2014}. Starting from $R^{'}_{HK}$ data and by using the Parker solar wind model, they studied the effect of stellar winds on the magnetospheric extension of fictitious Earth-like planets orbiting a sample of stars, including the subset of ten stars we selected for this case study. 
Considering that the typical error associated to the magnetosphere sizes computed by \cite{See2014} can be estimated as $\pm0.4$ $R_E$ from their Fig. 1, we can conclude that our results are in agreement within the confidence intervals.

%
%
\section{Discussion and Conclusions}
\label{S-Conclusions} 

In the present article we introduce a convenient relationship to deduce the long-term variability of the solar wind by studying its correlation with the solar magnetic activity. To this scope, we used an index that measures the chromospheric emission in the Ca II K resonance line.
The relationships between two solar wind parameters, speed and dynamic pressure, and the Ca II K index covering almost 5 solar cycles have been studied using 37-month averaged data.\\ 
Our study, based on cross-correlation techniques and mutual information analysis, shows that the solar wind properties follow the solar activity behaviour with a time lag of 3.2 years for the solar wind speed and 3.6 years for the dynamic pressure. The derived linear relationships between the Ca II K index and the solar wind speed and dynamic pressure are found to be valid for the whole time interval (five solar cycles) covered by this investigation.
\\
In Figure \ref{Shifted scatter plot} we showed the correlation between Ca II K index and solar wind parameters. Hysteresis cycles that could be linked to the 11-year solar cycles are made evident once the evolution in time is made explicit using the colour code. This is particularly evident in the case of the solar wind pressure, where parallel branches of the hysteresis cycle have the same slope of the fit representing the correlation between the parameters. Other cases of hysteresis cycles involving solar activity proxies are known \citep[see e.g.,][]{hyst2018MUPB...73..216B,hyst2019SoPh..294....8R,hyst2019SoPh..294...86S}. We plan to study the hysteresis cycles shown in this paper more in details in a future work, exploring the effects of a dynamic lag between the variables as opposed to the constant lag considered in this paper. A possible outcome of this future study is a new relation able to constrain the scatter plots on the fitted relation, providing higher correlation coefficients.
\\
Having a relationship that links the solar wind variations to that of Ca II K index is remarkable both for an historical reconstruction of the solar wind parameters and to even filling gaps for which measurements are not available. In particular, the solar Ca II K index is available since the beginning of the 20th century, but it has been synthetically reconstructed since 1750 in \citet{berrilli2020} by using different solar atmospheric models that represent quiet and magnetic regions. Furthermore, the time-shifted relations obtained can be employed for an attempt at short-time predictions into the future (up to 1.7 years for the solar wind speed and 2.1 years for its dynamic pressure). Those predictions could be very useful to assess the mean solar wind conditions from the point of view of human space missions, but also to forecast solar wind parameters during the flight phase of solar focused missions like Parker Solar Probe or Solar Orbiter, as in \cite{Venzmer2017}.
In particular our model predicts a minimum in the solar wind dynamic pressure in mid-2022. \\
We believe that these results are not only helpful to achieve a better knowledge of the interaction between Sun and Earth, but also for developing new skills to study the space-climate variability of other solar-type stars, in particular those with exoplanets, enabling us to characterise the interactions between planets and their host stars and the wind conditions of exoplanetary environment. Measurements in the Ca II H \& K lines are still performed by different ground-base telescopes, even directly expressed in terms of the S-index as at the TIGRE telescope \citep[see e.g.,][]{Schmitt2014, Gonzalez-Perez2022}. Since our relations have been calibrated on an intermediate age star, as the Sun, the most appropriate targets to which extend them are constituted by faculae dominated G-spectral type stars, characterised by an age $\gtrsim$ 2.6 Gyr and with a Rossby number $R_{0} \gtrsim 1$ \citep{Reinhold19}. For the future, we plan to compare the results from our model to that of other independent models, in order to assess the possibility to extend our procedure also to late F-type or early K-type stars. \\
In analogy to the case of the Sun, by making use of the relations we found, the variation and hence the effects of the stellar wind of Sun-like stars on their planets can be studied by analysing the temporal evolution of the chromospheric measurements already collected for several targets. Given the impossibility to obtain in-situ measurements of the stellar wind, it is difficult to recover the phase lag between stellar UV emission and stellar wind properties, and thus the stellar wind level in a precise moment in time. Nevertheless, having the information on at least a complete UV stellar cycle,
our model enables to compute the mean magnetospheric standoff distance for planets nearby Sun-like stars.\\
It is useful to point out that asteroseismic observations, like those obtained by the successful photometric space missions, i.e.,  {\it Kepler} \citep{Borucki2010} or TESS \citep{Ricker2014}, could also provide accurate fundamental parameters, in particular ages with an accuracy better than $15\%$ \citep[see e.g.,][]{Chaplin2014, Lebreton2014, DiMauro2017}, which can be used in combination with the present method to study the variation of the magnetospheric extension as a star like the Sun evolves, by comparing results for stars with different ages. 
In addition,  asteroseismic observations could also be used to study the variation of the magnetic activity of a Sun-like star by analysing the temporal evolution of oscillation parameters \citep[see e.g.,][]{Reda2022, DiMauro2022}.
In fact, it has been demonstrated by \cite{Bonanno2014} for a sample of 19 Sun-like stars, the presence of clear relations between the S-index and some asteroseismic parameters such as the amplitude of the observed acoustic oscillation modes or the 'small frequency separation', known as age indicator.
The targets selected in Table \ref{tab_magnetosphere} have been observed by TESS in 120~s and 20~s cadence mode and the asteroseismic analysis will be considered in the next future for a comparison with the present results.
Clearly, 
this  will open an independent way
to estimate the erosion of exoplanetary atmospheres.
Thus, we plan for the future to extend the present analysis to a wider set of stars, by exploiting the relationship between solar wind  properties and UV emission over the Mount Wilson Observatory (MWO) measurements, which regularly observed the Ca II H \& K emission since 1966 \citep{Wilson78} for several stars of different spectral types.
Further, in order to complete the phenomenological scenario, we will complement and verify our results by employing the independent procedure based on the asteroseismic method.

\section*{Acknowledgements}
 RR and PG are PhD students of the PhD course in Astronomy, Astrophysics and Space Science, a joint research program between the University of Rome “Tor Vergata”, the Sapienza University of Rome, and the National Institute of Astrophysics (INAF).
 The authors thank the reviewer P. Judge (HAO, NCAR) for the thoughtful comments that improved the quality of this paper.

\section*{Data Availability}

The Time series of the Ca II K index uses SOLIS data obtained by the NSO Integrated Synoptic Program (NISP), downloaded from the SOLIS website (\url{https://solis.nso.edu/0/iss/}; Monthly-Averaged ISS/SP/KKL Ca II K 1Å Emission Index Time Series). NISP is managed by the National Solar Observatory, which is operated by the Association of Universities for Research in Astronomy (AURA), Inc. under a cooperative agreement with the National Science Foundation. The Mg II composite is available from the University of Bremen (\url{http://www.iup.uni-bremen.de/UVSAT/Datasets/mgii}). The OMNI data are available from Coordinated Data AnalysisWeb (CDAWeb; \url{http://cdaweb.gsfc.nasa.gov}).
The datasets of the HK Project at the Mount Wilson Observatory are available from the National Solar Observatory (NSO) website (\url{https://nso.edu/data/historical-data/mount-wilson-observatory-hk-project/}).



\bibliographystyle{mnras}
\bibliography{biblio} 








\bsp	
\label{lastpage}


\end{document}